\documentclass[12pt,aps,english,prd,onecolumn,nofootinbib,superscriptaddress]{revtex4-2}

\usepackage[utf8]{inputenc}
\usepackage[english]{babel}  

\usepackage{textcomp}
\usepackage{amsmath,amssymb,amsfonts}
\usepackage{bm}
\usepackage{graphicx}
\usepackage{xcolor}               
\usepackage{color}                
\usepackage{textcomp}             
\usepackage{units}                
\usepackage[a4paper, margin=1.6cm]{geometry}
\usepackage{hyperref}             
\hypersetup{
    colorlinks=true,              
    breaklinks=true,              
    citecolor=blue,               
    linkcolor=[rgb]{0,0.5,0.9},   
    urlcolor=red,                 
    filecolor=green               
}

\newcommand{\orcid}[1]{\href{https://orcid.org/#1}{\includegraphics[width=10pt]{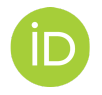}}}

\begin{document}

\title{Electromagnetic dynamics and geometric transport in spin-nondegenerate SME particles}

\author{A. A. Ara\'{u}jo Filho\orcid{0000-0002-8790-3944}}
\email{dilto@fisica.ufc.br}
\affiliation{Departamento de Física, Universidade Federal da Paraíba, Caixa Postal 5008, 58051--970, João Pessoa, Paraíba,  Brazil.}
\affiliation{Departamento de Física, Universidade Federal de Campina Grande Caixa Postal 10071, 58429-900 Campina Grande, Paraíba, Brazil.}
\affiliation{Center for Theoretical Physics, Khazar University, 41 Mehseti Street, Baku, AZ-1096, Azerbaijan.}
\author{A. F. Santos\orcid{0000-0002-2505-5273}}
\email{alesandroferreira@fisica.ufmt.br}
\affiliation{Programa de Pós-graduação em Física, Instituto de Física, Universidade Federal de Mato Grosso, Cuiabá, Brasil}
\author{J. A. A. S. Reis\orcid{0000-0002-2831-5317}}
\email{joao.reis@uesb.edu.br}
\affiliation{Departamento de Ciências Exatas e Naturais, Universidade Estadual do Sudoeste da Bahia, Campus Juvino Oliveira, Itapetinga -- BA, 45700-00, Brazil}
\author{L. Lisboa-Santos\orcid{0000-0003-4939-3856}}
\email{leticia.lisboa@discente.ufma.br}
\affiliation{Programa de Pós-Graduação em Física, Universidade Federal do Maranh\~{a}o, Campus Universit\'{a}rio do Bacanga, S\~{a}o Lu\'{i}s (MA), 65085-580, Brazil}
\author{V. B. Bezerra\orcid{0000-0001-7893-0265}}
\email{valdir@fisica.ufpb.br}
\affiliation{Departamento de Física, Universidade Federal da Paraíba, Caixa Postal 5008, 58051--970, João Pessoa, Paraíba,  Brazil.}

\date{\today}

\begin{abstract}

We investigate the electromagnetic dynamics of spin--nondegenerate classical particle models arising from Lorentz--violating sectors of the Standard--Model Extension, focusing on the $b_\mu$ background. Starting from the type--2 relativistic Lagrangian, we introduce minimal electromagnetic coupling and derive the exact Hamiltonian dynamics associated with each sector in terms of the gauge--covariant kinetic momentum. The modified dispersion relation leads to a sector--dependent relation between velocity and momentum, which directly affects the response to external fields. In the presence of a uniform magnetic field, we show that the two sectors exhibit distinct cyclotron frequencies and radii, implying that even constant fields dynamically resolve the underlying structure of the theory. In the nonrelativistic regime, the Lorentz--violating background induces a sector--dependent modification of the transverse inertial response, which can be interpreted as an effective anisotropic mass. After projection onto a single sector, the reduced dynamics acquires a noncanonical symplectic structure. The equations of motion can be written in semiclassical form with an effective momentum--space curvature $\Omega_{\pm}$, leading to anomalous velocity terms and a modified phase--space measure. As a consequence, a purely electric field generates opposite transverse drifts proportional to $q\,\mathbf{E} \times \Omega_{\pm}$, producing a Hall--like current without requiring a magnetic field.

\end{abstract}

\maketitle

\tableofcontents

\section{Introduction}

As it is well--known, the Lorentz invariance plays a central role in the formulation of both general relativity and the Standard Model. Nonetheless, several developments in high-energy physics and quantum gravity indicate that this symmetry may not be fundamental, but instead emerge as an effective feature at accessible energy scales. In particular, mechanisms such as spontaneous symmetry breaking, nontrivial vacuum configurations, and Planck-scale modifications provide consistent scenarios in which small departures from exact Lorentz invariance can arise \cite{KosteleckySamuel1989,ColladayKostelecky1997,ColladayKostelecky1998,Mattingly2005,AmelinoCamelia2013,Liberati2013,AraujoFilho:2024iox,araujo2025gravitational}.

A general and systematic description of these effects is provided by the Standard-Model Extension (SME), which parametrizes all Lorentz-violating operators compatible with observer covariance within an effective-field-theory framework \cite{ColladayKostelecky1998,Kostelecky2004,KosteleckyRussell2011,KosteleckyRussell2011}. This approach has been widely employed to investigate possible deviations from Lorentz symmetry in a broad range of physical systems, including particle interactions, precision spectroscopy, gravitational phenomena, and astrophysical observations \cite{KosteleckyMewes2008,KosteleckyTasson2011}.

In addition to the field-theoretic description, the classical limit of SME fermion sectors can be formulated in terms of relativistic point-particle models that reproduce the corresponding modified dispersion relations. These constructions establish a direct connection between Lorentz-violating operators and observable kinematical effects \cite{KosteleckyRussell2010,KosteleckyMewes2013,Kostelecky2011}. In several sectors, the dispersion relations exhibit spin-dependent splitting, leading to multiple dynamical branches. This structure admits a geometric interpretation in terms of generalized spacetime frameworks closely related to Finsler geometry, which naturally encode anisotropic propagation effects \cite{Kostelecky2011,PfeiferWohlfarth2012,GirelliLiberatiSindoni2007}.

A convenient description of these systems is provided by the so-called type--2 relativistic Lagrangians, which encode the branch structure of the modified dispersion relations in a compact and operational form. In this formulation, each sector defines an independent dynamical sector, enabling a direct treatment of the corresponding kinematics and interactions. This framework has been successfully employed in the analysis of classical particle motion, effective geometries, and gravitational couplings in Lorentz-violating scenarios \cite{KosteleckyRussell2010,KosteleckyMewes2013,Kostelecky2011,Kostelecky2011,KosteleckyEdwards2016,Kostelecky:2016kfm,AraujoFilho:2026zyt}.

On the other hand, the electromagnetic sector of spin-nondegenerate SME particle dynamics has received comparatively limited attention. Once minimal coupling to gauge fields is introduced, the dynamics becomes sensitive to the modified relation between velocity and momentum, providing a direct probe of the underlying dispersion structure. As a consequence, the interaction with electromagnetic fields reflects the presence of Lorentz-violating backgrounds in a nontrivial way, leading to dynamical effects that do not arise in the absence of charge \cite{KosteleckyLehnert2001,KosteleckyMewes2002,KosteleckyLane1999}.

Analogously, geometric structures in momentum space have become central in the description of transport phenomena across different areas of physics. In semiclassical electron dynamics, Berry curvature modifies the phase space structure, generating anomalous velocity terms and transverse transport effects \cite{Berry1984,SundaramNiu1999,XiaoChangNiu2010,Nagaosa2010,HasanKane2010,Armitage2018}. Similar geometric contributions can arise whenever the symplectic structure of phase space is effectively deformed, even in classical or semiclassical settings.

In Ref.~\cite{Reis:2026rch}, an alternative class of SME Lagrangians was constructed using an einbein, allowing a consistent treatment of both massive and massless particles. This formulation establishes a direct link between modified dispersion relations and classical dynamics, reproduces known fermion-sector results, and clarifies their classical content. A remarkable feature is the smooth massless limit, which makes the framework suitable for describing Lorentz-violating photon propagation. The constraint structure, handled via Dirac’s method, shows that the modified dispersion relations act as dynamical constraints. Based on this new methodology, in this work, we show that an analogous structure naturally emerges in the electromagnetic dynamics of spin-nondegenerate SME particles. Starting from the minimally coupled type--2 Lagrangian, we construct the Hamiltonian formulation in terms of gauge-covariant variables and derive the corresponding equations of motion. Although the kinetic momentum satisfies the standard Lorentz-force law, the relation between velocity and momentum becomes branch dependent, leading to observable dynamical effects. In the nonrelativistic regime, these features manifest in a clear manner. For constant Lorentz-violating backgrounds and uniform magnetic fields, the distinct branches exhibit different cyclotron frequencies and orbital radii. This implies that even homogeneous electromagnetic fields act as effective probes of the underlying dispersion structure. Furthermore, after projecting onto a single branch, the reduced phase space dynamics acquires a noncanonical symplectic structure that can be interpreted as an effective curvature in momentum space. The resulting equations of motion take a semiclassical form, exhibiting anomalous velocity contributions analogous to those found in Berry-curvature transport theory.

\section{Electromagnetic minimal coupling in the \(b_\mu\) sector}

We construct a framework for classical point-particle dynamics in Lorentz-violating backgrounds based on type-2 SME Lagrangians \cite{Reis:2026rch} with minimal electromagnetic coupling. Although equivalent to the standard square-root formulation after imposing constraints, this approach has a clearer structure: the canonical Hamiltonians are directly tied to the modified dispersion relations, which arise as secondary constraints in the Dirac--Bergmann procedure.

The construction is implemented for the $a_\mu$, $e_\mu$, $b_\mu$, and $H_{\mu\nu}$ sectors. In all cases, the type-2 Lagrangians reproduce the usual results after eliminating the einbein. The $a_\mu$ sector shifts momentum space, $e_\mu$ deforms the effective metric, $b_\mu$ yields a two-branch structure with spin dependence, and $H_{\mu\nu}$ leads to multi-branch dispersion relations.

As we shall see, an important feature is that the momentum can be inverted for the velocity even when the standard formulation fails, ensuring a consistent Hamiltonian description. The electromagnetic coupling is also naturally incorporated in the presence of the background fields.

In this section, we introduce the minimal electromagnetic coupling in the spin--nondegenerate $b_\mu$ sector and derive the corresponding equations of motion for each branch. Our purpose is to formulate the relativistic dynamics of a charged particle in a way that will be suitable for subsequent comparison with semiclassical transport and Berry--curvature--like structures.

Throughout this section, spacetime is assumed to be Minkowski, with metric $\eta_{\mu\nu}=\mathrm{diag}(+1,-1,-1,-1)$. We take the Lorentz-violating background $b_\mu$ to be constant, while the electromagnetic potential $A_\mu(x)$ remains arbitrary unless otherwise specified.

\subsection{Minimally coupled type--2 Lagrangian}

The type--2 Lagrangian for the spin--nondegenerate $b_\mu$ sector is constructed so as to encode, in a unified manner, the two independent branches of the modified dispersion relation associated with the Lorentz--violating background. In other words, this allows each branch to be treated as a distinct dynamical sector at the classical level
\begin{equation}
\tilde L_b^{(\pm)}
=
-\frac12
\left(
\frac{\dot x^2}{\mathfrak e}
\pm
2\sqrt{(b\cdot \dot x)^2-b^2\dot x^2}
+
\mathfrak e m^2
\right).
\label{eq:L-b-free}
\end{equation}
The electromagnetic minimal coupling is introduced by adding the standard
interaction term
\begin{equation}
L_{\rm em}=qA_\mu(x)\dot x^\mu.
\end{equation}
In this manner, the charged type--2 Lagrangian becomes
\begin{equation}
\tilde L_{b,\rm em}^{(\pm)}
=
-\frac12
\left(
\frac{\dot x^2}{\mathfrak e}
\pm
2\sqrt{(b\cdot \dot x)^2-b^2\dot x^2}
+
\mathfrak e m^2
\right)
+
qA_\mu(x)\dot x^\mu.
\label{eq:L-b-em}
\end{equation}

The variation with respect to the einbein remains unaffected, since the electromagnetic contribution does not depend on $\mathfrak{e}$. As a result, the associated constraint equation retains the same form and is given by
\begin{equation}
\frac{\partial \tilde L_{b,\rm em}^{(\pm)}}{\partial \mathfrak e}
=
\frac12\left(\frac{\dot x^2}{\mathfrak e^2}-m^2\right)=0,
\end{equation}
which simply implies
\begin{equation}
\dot x^2=\mathfrak e^2 m^2.
\label{eq:einbein-constraint-em}
\end{equation}
This condition enforces the mass--shell constraint and confirms that the
einbein $\mathfrak e$ remains a Lagrange multiplier, fixing the
normalization of the worldline parameter without being affected by the
electromagnetic coupling.

\subsection{Canonical and kinetic momenta}

We now introduce the canonical momentum associated with the worldline
coordinates. Adopting the sign convention
\begin{equation}
P_\mu:=-\frac{\partial \tilde L_{b,\rm em}^{(\pm)}}{\partial \dot x^\mu},
\end{equation}
we obtain
\begin{equation}
P_\mu
=
\frac{\dot x_\mu}{\mathfrak e}
\pm
\frac{(b\cdot\dot x)b_\mu-b^2\dot x_\mu}
{\sqrt{(b\cdot\dot x)^2-b^2\dot x^2}}
-qA_\mu(x).
\label{eq:canonical-P}
\end{equation}
The last term arises from the minimal electromagnetic coupling and
corresponds to the standard gauge contribution to the canonical
momentum.

It is convenient to separate this gauge--dependent piece by defining the
gauge--covariant (kinetic) momentum
\begin{equation}
\Pi_\mu := P_\mu + qA_\mu(x).
\label{eq:kinetic-momentum}
\end{equation}
In terms of $\Pi_\mu$, Eq.~\eqref{eq:canonical-P} becomes
\begin{equation}
\Pi_\mu
=
\frac{\dot x_\mu}{\mathfrak e}
\pm
\frac{(b\cdot\dot x)b_\mu-b^2\dot x_\mu}
{\sqrt{(b\cdot\dot x)^2-b^2\dot x^2}}.
\label{eq:Pi-velocity}
\end{equation}

This expression coincides exactly with the relation obtained in the absence of electromagnetic interactions. In this sense, the minimal coupling prescription acts only as a shift in the canonical momentum, while the relation between velocity and kinetic momentum remains unchanged and continues to exhibit the same branch--dependent structure.

This feature is particularly important: all Lorentz-violating effects associated with the $b_\mu$ background are entirely encoded in the nonlinear relation between $\Pi_\mu$ and $\dot x^\mu$, whereas the electromagnetic field enters only through the standard gauge shift. Consequently, once the dynamics is written in terms of the kinetic momentum, the structure of the equations of motion closely parallels the neutral case. In addition, we note that $\Pi_\mu$ transforms covariantly under gauge transformations, in contrast with $P_\mu$, and therefore provides the appropriate variable for describing the physical (observable) momentum of the particle. This distinction will play a central role in the Hamiltonian formulation and in the identification of semiclassical transport structures in the following sections.

\subsection{Branch dispersion relation}

The free $b_\mu$ sector is characterized by two distinct branches of the modified dispersion relation, given by
\begin{equation}
\mathcal D_b^{(\pm)}(p)
=
p^2-b^2-m^2
\pm
2\sqrt{(b\cdot p)^2-b^2p^2}.
\end{equation}
Each one defines an independent sector, which reflects the nontrivial momentum--space structure induced by the Lorentz--violating background.

In the presence of an electromagnetic field, minimal coupling is implemented through the substitution
\begin{equation}
p_\mu \;\rightarrow\; \Pi_\mu = P_\mu + qA_\mu(x),
\end{equation}
so that the dispersion relation becomes
\begin{equation}
\mathcal D_{b,\rm em}^{(\pm)}(x,P)
=
\Pi^2-b^2-m^2
\pm
2\sqrt{(b\cdot \Pi)^2-b^2\Pi^2}.
\label{eq:Db-em}
\end{equation}

The structure of the dispersion relation is preserved under this replacement: the electromagnetic interaction enters only through the kinetic momentum $\Pi_\mu$, while the branch splitting encoded in the square--root term remains unchanged. In particular, no additional mixing between them is introduced at this level.

The canonical Hamiltonian follows directly from the constraint and can be written as
\begin{equation}
\tilde H_{b,\rm em}^{(\pm)}
=
-\frac{\mathfrak e}{2}\,
\mathcal D_{b,\rm em}^{(\mp)}(x,P).
\label{eq:H-b-em}
\end{equation}
As in the neutral case, the sign labeling of the Hamiltonian is opposite to that of the corresponding dispersion branch, a consequence of the einbein enforcing the constraint $\mathcal D_{b,\rm em}^{(\pm)}=0$. It is worth pointing out that the electromagnetic field does not modify the functional form of the constraint, but only the variables on which it depends. This separation will be useful when comparing different dynamical regimes, since the Lorentz--violating contributions and the gauge effects remain clearly identifiable.
\subsection{Branch-resolved velocity}

Since $\mathcal D_{b,\rm em}^{(\pm)}$ depends on the canonical momentum only through the combination $\Pi_\mu = P_\mu + qA_\mu(x)$, the Hamilton equation for the coordinates takes the form
\begin{align}
\dot x^\mu
&=
-\frac{\partial \tilde H_{b,\rm em}^{(\pm)}}{\partial P_\mu}
=
\frac{\mathfrak e}{2}
\frac{\partial \mathcal D_{b,\rm em}^{(\mp)}}{\partial P_\mu}
=
\frac{\mathfrak e}{2}
\frac{\partial \mathcal D_{b,\rm em}^{(\mp)}}{\partial \Pi_\mu}.
\end{align}
This relation shows that the velocity is controlled entirely by the derivative of the dispersion function with respect to the kinetic momentum, with no explicit dependence on the gauge potential.

Performing the differentiation, we obtain
\begin{equation}
\dot x^\mu
=
\mathfrak e
\left[
\Pi^\mu
\pm
\frac{(b\cdot \Pi)b^\mu-b^2\Pi^\mu}
{\sqrt{(b\cdot \Pi)^2-b^2\Pi^2}}
\right].
\label{eq:velocity-Pi}
\end{equation}
For convenience, we introduce the shorthand
\begin{equation}
Q_b^\mu(\Pi)
:=
\frac{(b\cdot \Pi)b^\mu-b^2\Pi^\mu}
{\sqrt{(b\cdot \Pi)^2-b^2\Pi^2}},
\label{eq:QbPi}
\end{equation}
so that the velocity can be written compactly as
\begin{equation}
\dot x^\mu
=
\mathfrak e\left(\Pi^\mu\pm Q_b^\mu(\Pi)\right).
\label{eq:velocity-compact}
\end{equation}

This expression makes explicit that the relation between velocity and kinetic momentum is nonlinear and branch dependent. The vector $Q_b^\mu(\Pi)$ is orthogonal to $\Pi^\mu$ and encodes the deformation induced by the $b_\mu$ background, thereby controlling the departure from the standard relativistic relation. In particular, the velocity is not aligned with $\Pi^\mu$, even in the presence of a uniform electromagnetic field. This misalignment is an intrinsic feature of the Lorentz--violating sector and persists independently of the gauge interaction. As a consequence, the direction of motion and the direction of momentum define distinct geometric structures in phase space, a property that will play an important role in
the subsequent dynamical analysis.

\subsection{Equation of motion for the kinetic momentum}

The canonical Hamilton equation for $P_\mu$ is
\begin{equation}
\dot P_\mu
=
\frac{\partial \tilde H_{b,\rm em}^{(\pm)}}{\partial x^\mu}.
\end{equation}
Since the background $b_\mu$ is constant, the Hamiltonian depends on the spacetime coordinates only through the electromagnetic potential $A_\mu(x)$. It is therefore convenient to express the evolution in terms of the kinetic momentum
\begin{equation}
\Pi_\mu = P_\mu + qA_\mu(x).
\end{equation}
Taking the derivative with respect to the worldline parameter, we find
\begin{equation}
\dot \Pi_\mu
=
\dot P_\mu
+
q\,\partial_\nu A_\mu\,\dot x^\nu.
\label{eq:Pi-dot-1}
\end{equation}

Evaluating $\dot P_\mu$ from the Hamiltonian and combining terms, one arrives at
\begin{equation}
\dot \Pi_\mu
=
qF_{\mu\nu}\dot x^\nu,
\label{eq:Pi-dot}
\end{equation}
where
\begin{equation}
F_{\mu\nu}:=\partial_\mu A_\nu-\partial_\nu A_\mu
\end{equation}
is the electromagnetic field strength tensor.

This equation has the same structural form as the relativistic Lorentz force law, now written in terms of the gauge--covariant momentum $\Pi_\mu$. In particular, the evolution depends only on the field strength tensor, making gauge invariance manifest at the level of the equations of motion.  The distinctive feature of the present framework lies in the relation between $\dot x^\mu$ and $\Pi^\mu$. Substituting Eq.~\eqref{eq:velocity-compact} into Eq.~\eqref{eq:Pi-dot}, one sees that the acceleration is governed by a velocity that is not collinear with the kinetic momentum and depends explicitly on the branch. As a result, the effective force experienced by the particle acquires a nontrivial dependence on the orientation of $\Pi_\mu$ relative to $b_\mu$.  This leads to a qualitatively modified dynamics: even in simple field configurations, the trajectories associated with the two branches differ.

\subsection{Three-vector form}

To make contact with the standard relativistic formulation, we rewrite the equations in three-vector notation. We decompose
\begin{equation}
\Pi^\mu=(\Pi^0,\boldsymbol{\Pi}),
\qquad
b^\mu=(b^0,\mathbf b),
\qquad
\dot x^\mu=(\dot t,\dot{\mathbf x}).
\end{equation}
In these variables, Eq.~\eqref{eq:Pi-dot} becomes
\begin{subequations}
\begin{align}
\dot \Pi^0
&=
q\,\mathbf E\cdot \dot{\mathbf x},
\\[1ex]
\dot{\boldsymbol{\Pi}}
&=
q\left(\mathbf E\,\dot t+\dot{\mathbf x}\times \mathbf B\right),
\end{align}
\end{subequations}
where
\begin{equation}
E_i:=F_{i0},
\qquad
B^i:=\frac12 \epsilon^{ijk}F_{jk}.
\end{equation}

The corresponding expressions for the velocity follow directly from
Eq.~\eqref{eq:velocity-compact}:
\begin{subequations}
\begin{align}
\dot t
&=
\mathfrak e\left(\Pi^0\pm Q_b^0\right),
\\[1ex]
\dot{\mathbf x}
&=
\mathfrak e\left(\boldsymbol{\Pi}\pm \mathbf Q_b\right),
\end{align}
\end{subequations}
with $Q_b^\mu$ defined in Eq.~\eqref{eq:QbPi}. Substituting these expressions into the momentum equation yields
\begin{equation}
\dot{\boldsymbol{\Pi}}
=
q\left[
\mathbf E\,\mathfrak e\left(\Pi^0\pm Q_b^0\right)
+
\mathfrak e\left(\boldsymbol{\Pi}\pm \mathbf Q_b\right)\times \mathbf B
\right].
\label{eq:Pi-dot-3vec}
\end{equation}

This form makes explicit that the electromagnetic response is governed by a velocity that differs from the kinetic momentum by the $b_\mu$-dependent contribution $\mathbf Q_b$. As a consequence, even when the field configuration is fixed, the evolution of $\boldsymbol{\Pi}$ depends on the branch through both temporal and spatial components of the motion.

\subsection{Constant magnetic field}

A particularly transparent regime is obtained for a purely magnetic configuration,
\begin{equation}
\mathbf E=0,
\qquad
\mathbf B=\text{const}.
\end{equation}
In this case, the dynamics reduces to
\begin{equation}
\dot{\boldsymbol{\Pi}}
=
q\,\mathfrak e\left(\boldsymbol{\Pi}\pm \mathbf Q_b\right)\times \mathbf B,
\label{eq:Pi-dot-B}
\end{equation}
together with
\begin{equation}
\dot{\mathbf x}
=
\mathfrak e\left(\boldsymbol{\Pi}\pm \mathbf Q_b\right).
\label{eq:x-dot-B}
\end{equation}

These relations show that the motion in a magnetic field is governed by an effective velocity that depends on the branch through $\mathbf Q_b$. Even when $\boldsymbol{\Pi}$ is fixed, the two branches correspond to different directions and magnitudes of $\dot{\mathbf x}$, which directly affects the magnetic force. As a result, the resulting trajectories are not universal: the circular motion typically associated with a constant magnetic field is modified in a sector--dependent way, leading to distinct cyclotron frequencies and orbital radii for the two sectors.


\subsection{Purely spacelike constant $b^\mu$}

In order to make the structure more transparent, we specialize to a purely spacelike and constant background,
\begin{equation}
b^\mu=(0,\mathbf b),
\qquad
\mathbf b=\text{const}.
\label{eq:b-spacelike-const}
\end{equation}
In this case,
\begin{equation}
b^2=-\mathbf b^2,
\qquad
b\cdot \Pi=-\,\mathbf b\cdot \boldsymbol{\Pi},
\end{equation}
and the Gramian entering the square--root term becomes
\begin{align}
(b\cdot \Pi)^2-b^2\Pi^2
&=
(\mathbf b\cdot \boldsymbol{\Pi})^2
+
\mathbf b^2\left[(\Pi^0)^2-\boldsymbol{\Pi}^2\right]
\notag\\
&=
\mathbf b^2(\Pi^0)^2
-
\mathbf b^2\boldsymbol{\Pi}^2
+
(\mathbf b\cdot \boldsymbol{\Pi})^2.
\label{eq:Gramian-spacelike-B}
\end{align}
Here, we have explicitly that the deformation depends not only on the magnitude of $\boldsymbol{\Pi}$, but also on its orientation relative to $\mathbf b$. In addition, the components of the branch deformation vector are then
\begin{subequations}
\begin{align}
Q_b^0
&=
\frac{\mathbf b^2\,\Pi^0}
{\sqrt{\mathbf b^2(\Pi^0)^2-\mathbf b^2\boldsymbol{\Pi}^2+(\mathbf b\cdot \boldsymbol{\Pi})^2}},
\\[1ex]
\mathbf Q_b
&=
\frac{
-(\mathbf b\cdot \boldsymbol{\Pi})\,\mathbf b+\mathbf b^2\boldsymbol{\Pi}
}{
\sqrt{\mathbf b^2(\Pi^0)^2-\mathbf b^2\boldsymbol{\Pi}^2+(\mathbf b\cdot \boldsymbol{\Pi})^2}
}.
\end{align}
\label{eq:Qb-spacelike}
\end{subequations}

Substituting into the velocity expression, we obtain
\begin{equation}
\dot{\mathbf x}
=
\mathfrak e
\left[
\boldsymbol{\Pi}
\pm
\frac{
-(\mathbf b\cdot \boldsymbol{\Pi})\,\mathbf b+\mathbf b^2\boldsymbol{\Pi}
}{
\sqrt{\mathbf b^2(\Pi^0)^2-\mathbf b^2\boldsymbol{\Pi}^2+(\mathbf b\cdot \boldsymbol{\Pi})^2}
}
\right].
\label{eq:x-dot-spacelike}
\end{equation}
Even in this simple configuration, the velocity depends nonlinearly on
$\boldsymbol{\Pi}$ and acquires an anisotropic contribution controlled by
$\mathbf b$. In particular, the deviation from alignment between velocity
and momentum is sensitive to the projection of $\boldsymbol{\Pi}$ along
the preferred direction $\mathbf b$.

\subsection{Immediate physical consequences}

Equations~\eqref{eq:Pi-dot-B} and \eqref{eq:x-dot-spacelike} already encode several direct consequences for the dynamics. However, the motion in a magnetic field is no longer characterized by a universal relation between momentum and velocity. For a fixed $\boldsymbol{\Pi}$, the two sectors correspond to different velocities, which implies that the effective magnetic force differs between them. As a result, the associated cyclotron motion is branch dependent, leading to distinct orbital radii and frequencies.

Moreover, since the velocity depends on the orientation of $\boldsymbol{\Pi}$ relative to $\mathbf b$, the center of the orbital motion is also affected. Even when the initial kinetic momentum is the same, the trajectories associated with each branch do not share the same guiding center.

Another important feature is that the map \begin{equation} \boldsymbol{\Pi}\mapsto \dot{\mathbf x}_\pm(\boldsymbol{\Pi}) \end{equation} is nonlinear. This implies that momentum space acquires a nontrivial structure, in the sense that the velocity cannot be obtained from a simple proportionality relation. This property will be relevant when formulating the dynamics in terms of effective phase--space variables.

Nevertheless, after the inclusion of electromagnetic coupling, the system is governed by the gauge--covariant dispersion relations
\begin{equation}
\mathcal D_{b,\rm em}^{(\pm)}(x,P)
=
\Pi^2-b^2-m^2
\pm
2\sqrt{(b\cdot \Pi)^2-b^2\Pi^2},
\qquad
\Pi_\mu=P_\mu+qA_\mu,
\end{equation}
together with the evolution equation
\begin{equation}
\dot \Pi_\mu=qF_{\mu\nu}\dot x^\nu.
\end{equation}
The nontrivial aspect of the dynamics is entirely encoded in the branch--dependent relation between $\dot x^\mu$ and $\Pi^\mu$, which persists even for constant $b_\mu$ and leads to distinct trajectories under identical external conditions.

\section{Nonrelativistic limit in a constant magnetic field}

In this section, we derive the nonrelativistic limit of the minimally coupled $b_\mu$ sector in the presence of a constant magnetic field. Our main purpose is to obtain an explicit branch--dependent cyclotron dynamics and to exhibit, in a concrete setting, the first dynamical consequences of the nontrivial momentum dependence induced by the spin--nondegenerate background.

We continue to work in Minkowski spacetime and assume a constant purely magnetic external field,
\begin{equation}
\mathbf E=0,
\qquad
\mathbf B=\mathrm{const},
\end{equation}
together with a constant purely spacelike Lorentz--violating background,
\begin{equation}
b^\mu=(0,\mathbf b),
\qquad
\mathbf b=\mathrm{const},
\end{equation}
and the slow--motion regime
\begin{equation}
|\boldsymbol{\Pi}|\ll m,
\end{equation}
where $m$ denotes the fermion mass.

\subsection{Relativistic branch velocity revisited}

From the previous section, the exact branch--resolved velocity is
\begin{equation}
\dot{\mathbf x}
=
\mathfrak e
\left(
\boldsymbol{\Pi}\pm \mathbf Q_b
\right),
\label{eq:xdot-exact-NRB}
\end{equation}
where
\begin{equation}
\mathbf Q_b
=
\frac{
-(\mathbf b\cdot \boldsymbol{\Pi})\,\mathbf b+\mathbf b^2\boldsymbol{\Pi}
}{
\sqrt{
\mathbf b^2(\Pi^0)^2-\mathbf b^2\boldsymbol{\Pi}^2+(\mathbf b\cdot \boldsymbol{\Pi})^2
}
}.
\label{eq:Qb-exact-NRB}
\end{equation}
At the same time, in the purely magnetic case the kinetic momentum obeys
\begin{equation}
\dot{\boldsymbol{\Pi}}
=
q\,\dot{\mathbf x}\times \mathbf B.
\label{eq:Pidot-exact-NRB}
\end{equation}

The nonrelativistic limit is obtained by expanding the velocity for small
$|\boldsymbol{\Pi}|/m$. This will allow us to identify the leading
branch-dependent correction to the standard magnetic dynamics.

\subsection{Nonrelativistic expansion of the branch velocity}

Let us write
\begin{equation}
\Pi^0 = m + \varepsilon,
\qquad
|\varepsilon|\ll m.
\end{equation}
Then the denominator of Eq.~\eqref{eq:Qb-exact-NRB} becomes
\begin{align}
\sqrt{
\mathbf b^2(\Pi^0)^2-\mathbf b^2\boldsymbol{\Pi}^2+(\mathbf b\cdot \boldsymbol{\Pi})^2
}
&=
|\mathbf b|\,
\sqrt{
(\Pi^0)^2-\boldsymbol{\Pi}^2+\frac{(\mathbf b\cdot \boldsymbol{\Pi})^2}{\mathbf b^2}
}
\notag\\
&\approx
|\mathbf b|\,m
\left[
1
+
\frac{1}{2m^2}
\left(
2m\varepsilon-\boldsymbol{\Pi}^2+\frac{(\mathbf b\cdot \boldsymbol{\Pi})^2}{\mathbf b^2}
\right)
\right].
\label{eq:denominator-expand}
\end{align}
To leading order, it is therefore sufficient to keep
\begin{equation}
\sqrt{
\mathbf b^2(\Pi^0)^2-\mathbf b^2\boldsymbol{\Pi}^2+(\mathbf b\cdot \boldsymbol{\Pi})^2
}
\approx
|\mathbf b|\,m.
\end{equation}
Hence,
\begin{equation}
\mathbf Q_b
\approx
\frac{1}{m}
\left[
|\mathbf b|\,\boldsymbol{\Pi}
-
\frac{\mathbf b\cdot \boldsymbol{\Pi}}{|\mathbf b|}\,\mathbf b
\right].
\label{eq:Qb-leading}
\end{equation}

Introducing the unit vector
\begin{equation}
\hat{\mathbf b}:=\frac{\mathbf b}{|\mathbf b|},
\end{equation}
Eq.~\eqref{eq:Qb-leading} can be rewritten as
\begin{equation}
\mathbf Q_b
\approx
\frac{|\mathbf b|}{m}
\left[
\boldsymbol{\Pi}
-(\hat{\mathbf b}\cdot \boldsymbol{\Pi})\,\hat{\mathbf b}
\right].
\label{eq:Qb-projected}
\end{equation}
This makes clear that only the component of $\boldsymbol{\Pi}$ transverse to $\hat{\mathbf b}$ contributes to the branch correction.

It is then convenient to decompose the kinetic momentum into parallel and orthogonal components relative to $\hat{\mathbf b}$:
\begin{equation}
\boldsymbol{\Pi}
=
\boldsymbol{\Pi}_\parallel+\boldsymbol{\Pi}_\perp,
\qquad
\boldsymbol{\Pi}_\parallel:=(\hat{\mathbf b}\cdot \boldsymbol{\Pi})\hat{\mathbf b},
\qquad
\boldsymbol{\Pi}_\perp:=\boldsymbol{\Pi}-(\hat{\mathbf b}\cdot \boldsymbol{\Pi})\hat{\mathbf b}.
\end{equation}
Then Eq.~\eqref{eq:Qb-projected} becomes simply
\begin{equation}
\mathbf Q_b \approx \frac{|\mathbf b|}{m}\,\boldsymbol{\Pi}_\perp.
\end{equation}

Therefore, the velocity reads
\begin{equation}
\dot{\mathbf x}
\approx
\mathfrak e
\left(
\boldsymbol{\Pi}
\pm
\frac{|\mathbf b|}{m}\,\boldsymbol{\Pi}_\perp
\right).
\label{eq:velocity-expand-1}
\end{equation}

To recover the conventional nonrelativistic normalization, we use the
einbein constraint $\dot x^2=\mathfrak e^2m^2$, which in the present
regime implies
\begin{equation}
\mathfrak e \approx \frac{1}{m}.
\label{eq:einbein-NR}
\end{equation}
Therefore,
\begin{equation}
\dot{\mathbf x}_\pm
\approx
\frac{\boldsymbol{\Pi}}{m}
\pm
\frac{|\mathbf b|}{m^2}\,\boldsymbol{\Pi}_\perp.
\label{eq:vpm-NR}
\end{equation}

Equation~\eqref{eq:vpm-NR} gives the leading nonrelativistic, branch--resolved velocity in the magnetic configuration under consideration.

\subsection{Effective mass anisotropy}

Equation~\eqref{eq:vpm-NR} admits a simple interpretation. Along the direction parallel to $\mathbf b$,
\begin{equation}
\dot{\mathbf x}_{\parallel,\pm}
=
\frac{\boldsymbol{\Pi}_\parallel}{m},
\end{equation}
so the motion remains unchanged at leading order. By contrast, in the transverse plane we find
\begin{equation}
\dot{\mathbf x}_{\perp,\pm}
=
\left(
\frac{1}{m}
\pm
\frac{|\mathbf b|}{m^2}
\right)\boldsymbol{\Pi}_\perp.
\end{equation}
The Lorentz--violating background therefore modifies only the transverse response and does so in a branch--dependent manner. This can be expressed as an effective transverse mass renormalization:
\begin{equation}
\frac{1}{m_{\perp,\pm}^{\rm eff}}
=
\frac{1}{m}
\pm
\frac{|\mathbf b|}{m^2}.
\label{eq:meff-inverse}
\end{equation}
Equivalently,
\begin{equation}
m_{\perp,\pm}^{\rm eff}
\approx
m\left(1\mp \frac{|\mathbf b|}{m}\right).
\label{eq:meff-direct}
\end{equation}

In other words, the two sectors couple differently to the magnetic field because
their transverse inertial response is not the same.

\subsection{Branch-dependent cyclotron dynamics}

Substituting Eq.~\eqref{eq:vpm-NR} into the magnetic force law
\eqref{eq:Pidot-exact-NRB}, we obtain
\begin{equation}
\dot{\boldsymbol{\Pi}}
=
q\left[
\frac{\boldsymbol{\Pi}}{m}
\pm
\frac{|\mathbf b|}{m^2}\,\boldsymbol{\Pi}_\perp
\right]\times \mathbf B.
\label{eq:Pidot-NR-1}
\end{equation}

To extract the clearest cyclotron picture, let us specialize to the case
\begin{equation}
\mathbf B \parallel \mathbf b.
\label{eq:Bparallelb}
\end{equation}
Here, the transverse projection with respect to $\mathbf b$ coincides with the transverse projection with respect to $\mathbf B$, so that only the transverse momentum undergoes rotation. Writing
\begin{equation}
\mathbf B = B\,\hat{\mathbf z},
\qquad
\mathbf b = |\mathbf b|\,\hat{\mathbf z},
\end{equation}
one finds
\begin{equation}
\dot{\boldsymbol{\Pi}}_\parallel = 0,
\end{equation}
and
\begin{equation}
\dot{\boldsymbol{\Pi}}_\perp
=
q
\left(
\frac{1}{m}\pm \frac{|\mathbf b|}{m^2}
\right)
\boldsymbol{\Pi}_\perp \times \mathbf B.
\label{eq:Pi-perp-rotate}
\end{equation}
This is precisely uniform cyclotron rotation with branch--dependent frequency
\begin{equation}
\omega_{c,\pm}
=
\frac{|q|B}{m}
\left(
1\pm \frac{|\mathbf b|}{m}
\right).
\label{eq:omega-c-pm}
\end{equation}
Therefore,
\begin{equation}
\omega_{c,+}\neq \omega_{c,-}.
\end{equation}

The corresponding cyclotron radii are
\begin{equation}
r_{c,\pm}
=
\frac{|\boldsymbol{\Pi}_\perp|}{|q|B}
\left(
1\mp \frac{|\mathbf b|}{m}
\right)
+\mathcal O\!\left(\frac{|\mathbf b|^2}{m^2}\right),
\label{eq:rc-pm}
\end{equation}
so that
\begin{equation}
r_{c,+}\neq r_{c,-}.
\end{equation}

A uniform magnetic field therefore already resolves the two branches at the level of the orbital motion, even for a constant Lorentz--violating background.

Equations~\eqref{eq:omega-c-pm} and \eqref{eq:rc-pm} show that a constant magnetic field probes the branch structure of the Lorentz--violating particle model in a direct way. Even without gradients of $\mathbf b$, the two branches display different orbital responses because the velocity--momentum relation is itself branch dependent.  In this sense, the magnetic field acts as a dynamical branch resolver: the frequencies differ, the orbital radii differ, and the two motions separate in time under identical external conditions. At the nonrelativistic level, this behavior resembles the cyclotron response of modes with distinct transverse inertial parameters, although here the splitting originates from the relativistic branch structure of the underlying theory.

\subsection{Solution of the transverse motion}

For $\mathbf B\parallel \mathbf b$, the transverse momentum satisfies
\begin{equation}
\frac{\mathrm d\boldsymbol{\Pi}_\perp}{\mathrm dt}
=
\omega_{c,\pm}\,\boldsymbol{\Pi}_\perp\times \hat{\mathbf z},
\end{equation}
whose solution is
\begin{equation}
\boldsymbol{\Pi}_{\perp,\pm}(t)
=
\Pi_\perp
\left(
\cos \omega_{c,\pm} t\,\hat{\mathbf x}
+
\sin \omega_{c,\pm} t\,\hat{\mathbf y}
\right),
\label{eq:Pi-solution}
\end{equation}
up to an initial phase fixed by the chosen initial conditions. The corresponding transverse velocity is
\begin{equation}
\dot{\mathbf x}_{\perp,\pm}
=
\left(
\frac{1}{m}\pm \frac{|\mathbf b|}{m^2}
\right)
\boldsymbol{\Pi}_{\perp,\pm}(t).
\end{equation}
Then, the transverse trajectories are circles of radius $r_{c,\pm}$ traversed with frequency $\omega_{c,\pm}$. The parallel motion remains
uniform,
\begin{equation}
\dot{\mathbf x}_{\parallel,\pm}=\frac{\boldsymbol{\Pi}_\parallel}{m},
\end{equation}
so that the full trajectory is a branch dependent helix.

\subsection{Branch splitting in the time domain}

Suppose that the two branches are launched with the same initial kinetic
momentum $\boldsymbol{\Pi}_\perp(0)$. Since their cyclotron frequencies
are different, a relative phase accumulates in time:
\begin{equation}
\Delta\phi(t)
=
(\omega_{c,+}-\omega_{c,-})t
=
2\,\frac{|q|B\,|\mathbf b|}{m^2}\,t.
\label{eq:Delta-phase}
\end{equation}
Even in uniform fields, the trajectories progressively separate in
configuration space.

The branch difference in cyclotron frequency is
\begin{equation}
\Delta\omega_c
:=
\omega_{c,+}-\omega_{c,-}
=
2\,\frac{|q|B\,|\mathbf b|}{m^2}.
\label{eq:Delta-omega}
\end{equation}
Likewise, the difference in cyclotron radius is
\begin{equation}
\Delta r_c
:=
r_{c,+}-r_{c,-}
=
-\,2\,\frac{|\boldsymbol{\Pi}_\perp|\,|\mathbf b|}{|q|B\,m}.
\label{eq:Delta-r}
\end{equation}

These relations make explicit how the Lorentz--violating background
induces a measurable splitting between the two sectors already at leading
nonrelativistic order.

\subsection{Toward a Berry-curvature-like interpretation}

The main lesson of the present analysis is that the branch velocity is a nonlinear function of the kinetic momentum,
\begin{equation}
\dot{\mathbf x}_\pm
=
\mathbf v_\pm(\boldsymbol{\Pi}),
\qquad
\mathbf v_\pm \neq \frac{\boldsymbol{\Pi}}{m}.
\end{equation}
Such a nontrivial momentum--velocity map is precisely the type of structure that, after projection onto a single branch, gives rise to an effective noncanonical phase--space description.

In the present relativistic particle model, the deformation $\mathbf Q_b(\boldsymbol{\Pi})$ retains the information associated with the eliminated branch and modifies the kinematics already before any
explicit Berry--curvature--like formulation is introduced. The branch dependence of $\omega_{c,\pm}$ and $r_{c,\pm}$ is one concrete manifestation of this fact.

The full Berry--curvature--like description will be developed later. At this stage, the nonrelativistic magnetic dynamics already shows that the minimally coupled $b_\mu$ model carries a nontrivial momentum space structure, visible through the splitting of the orbital motion. In the nonrelativistic limit, a constant purely spacelike $b_\mu$ background modifies the branch resolved velocity according to
\begin{equation}
\dot{\mathbf x}_\pm
\approx
\frac{\boldsymbol{\Pi}}{m}
\pm
\frac{|\mathbf b|}{m^2}\,\boldsymbol{\Pi}_\perp.
\end{equation}
This induces a sector--dependent effective transverse mass,
\begin{equation}
m_{\perp,\pm}^{\rm eff}
\approx
m\left(1\mp \frac{|\mathbf b|}{m}\right),
\end{equation}
and, in a constant magnetic field parallel to $\mathbf b$, a branch--dependent cyclotron motion with
\begin{equation}
\omega_{c,\pm}
=
\frac{|q|B}{m}
\left(1\pm \frac{|\mathbf b|}{m}\right),
\qquad
r_{c,\pm}
=
\frac{|\boldsymbol{\Pi}_\perp|}{|q|B}
\left(1\mp \frac{|\mathbf b|}{m}\right).
\end{equation}
Therefore, even a uniform magnetic field is sufficient to split the orbital motion of the two branches, revealing the nontrivial kinematics of the minimally coupled spin--nondegenerate SME particle model.

\section{Effective branch-projected symplectic structure and
Berry-curvature-like dynamics}

In the previous section we showed that, after electromagnetic minimal coupling, the velocity of the charged $b_\mu$ particle is not collinear with the kinetic momentum. In the nonrelativistic regime and for a constant purely spacelike background $b^\mu=(0,\mathbf b)$, the velocity takes the form
\begin{equation}
\dot{\mathbf x}_\pm
=
\frac{\boldsymbol{\Pi}}{m}
\pm
\frac{|\mathbf b|}{m^2}\,\boldsymbol{\Pi}_\perp
+\cdots,
\label{eq:vpm-recall}
\end{equation}
where $\boldsymbol{\Pi}_\perp$ denotes the component of $\boldsymbol{\Pi}$ orthogonal to $\mathbf b$. This already signals that,
after selecting a single branch, the reduced dynamics cannot be generated by the standard canonical phase space structure.

The goal of this section is to show that the one--branch dynamics admits a reformulation analogous to semiclassical transport with a nontrivial momentum space structure. The derivation is purely structural: the effective curvature introduced below does not rely on Bloch states, but organizes the projected SME dynamics in the same mathematical form.

\subsection{From the full branch dynamics to an effective reduced theory}

Let the charged branch Hamiltonian be
\begin{equation}
\tilde H_{b,\rm em}^{(\pm)}
=
-\frac{\mathfrak e}{2}\,
\mathcal D_{b,\rm em}^{(\mp)}(x,P),
\end{equation}
with
\begin{equation}
\Pi_\mu=P_\mu+qA_\mu(x).
\end{equation}
In the full phase space $(x^\mu,P_\mu)$, the Poisson structure is canonical. Once a branch is selected, however, the dynamics becomes restricted to the shell
\begin{equation}
\mathcal D_{b,\rm em}^{(\pm)}(x,P)=0,
\label{eq:branch-shell}
\end{equation}
and the eliminated branch is no longer dynamically accessible.

This restriction modifies the effective phase space description. In particular, the remaining degrees of freedom inherit a nontrivial momentum dependence, which must be encoded in the reduced symplectic structure.

\subsection{Effective branch Hamiltonian}

From the nonrelativistic analysis, the branch energies are
\begin{equation}
\varepsilon_\pm(\boldsymbol{\Pi})
=
\frac{\boldsymbol{\Pi}^2}{2m}
\mp
|\mathbf b|
\mp
\frac{|\mathbf b|}{2m^2}\,\boldsymbol{\Pi}_\perp^2
+\cdots,
\label{eq:eps-branch-NR}
\end{equation}
where higher--order terms in $|\mathbf b|/m$ have been omitted. Writing
\begin{equation}
\boldsymbol{\Pi}_\perp^2
=
\boldsymbol{\Pi}^2-(\hat{\mathbf b}\cdot\boldsymbol{\Pi})^2,
\qquad
\hat{\mathbf b}:=\frac{\mathbf b}{|\mathbf b|},
\end{equation}
the effective one--branch Hamiltonian becomes
\begin{equation}
H_{\rm eff,\pm}(\mathbf x,\boldsymbol{\Pi})
=
\varepsilon_\pm(\boldsymbol{\Pi})+q\phi(\mathbf x).
\label{eq:Heff-branch}
\end{equation}

The associated group velocity is
\begin{equation}
\mathbf v_\pm(\boldsymbol{\Pi})
=
\nabla_{\boldsymbol{\Pi}}\varepsilon_\pm(\boldsymbol{\Pi})
=
\frac{\boldsymbol{\Pi}}{m}
\pm
\frac{|\mathbf b|}{m^2}\,\boldsymbol{\Pi}_\perp
+\cdots,
\label{eq:v-from-eps}
\end{equation}
which reproduces Eq.~\eqref{eq:vpm-recall}.

\subsection{Modified symplectic structure and effective curvature}

If one were to retain a canonical symplectic structure, the resulting equations would reproduce the previously derived cyclotron motion. However, this description obscures the geometric content introduced by branch projection.

A more intrinsic formulation is obtained by allowing for a momentum space correction to the symplectic structure. At leading order, the equations of motion may be written as
\begin{subequations}
\begin{align}
\dot{\mathbf x}_\pm
&=
\nabla_{\boldsymbol{\Pi}}\varepsilon_\pm
-
\dot{\boldsymbol{\Pi}}_\pm\times \boldsymbol{\Omega}_\pm(\boldsymbol{\Pi}),
\\[1ex]
\dot{\boldsymbol{\Pi}}_\pm
&=
q\left(\mathbf E+\dot{\mathbf x}_\pm\times \mathbf B\right)
-\nabla_{\mathbf x}\varepsilon_\pm.
\end{align}
\label{eq:Berry-like-EOM}
\end{subequations}

The effective curvature $\boldsymbol{\Omega}_\pm(\boldsymbol{\Pi})$
encodes the momentum dependence inherited from the eliminated branch. It is not introduced from microscopic states, but emerges as the quantity that organizes the reduced dynamics in a closed form.

\subsection{Leading-order structure of the curvature}

The background selects a preferred direction $\hat{\mathbf b}$, so the curvature must be constructed from $\boldsymbol{\Pi}$ and
$\hat{\mathbf b}$. To leading order, the simplest consistent structure is
\begin{equation}
\boldsymbol{\Omega}_\pm
=
\pm \Omega_b\,\hat{\mathbf b}
+
\mathcal O\!\left(\frac{|\boldsymbol{\Pi}|}{m^2}\right),
\label{eq:Omega-leading}
\end{equation}
with
\begin{equation}
\Omega_b \sim \frac{|\mathbf b|}{m^2}.
\end{equation}

This form is sufficient to take into account the dominant correction to the sector-projected dynamics.

\subsection{Modified equations and cyclotron response}

In the purely magnetic case with $\mathbf E=0$, the equations reduce to
\begin{subequations}
\begin{align}
\dot{\mathbf x}_\pm
&=
\nabla_{\boldsymbol{\Pi}}\varepsilon_\pm
-
\dot{\boldsymbol{\Pi}}_\pm\times(\pm \Omega_b\hat{\mathbf b}),
\\[1ex]
\dot{\boldsymbol{\Pi}}_\pm
&=
q\,\dot{\mathbf x}_\pm\times \mathbf B.
\end{align}
\end{subequations}

Eliminating $\dot{\boldsymbol{\Pi}}_\pm$ and taking
$\mathbf B\parallel\hat{\mathbf b}$, one obtains
\begin{equation}
\dot{\mathbf x}_{\perp,\pm}
=
\frac{\nabla_{\boldsymbol{\Pi}_\perp}\varepsilon_\pm}
{1\mp qB\Omega_b}.
\label{eq:vperp-renormalized}
\end{equation}
Expanding to first order,
\begin{equation}
\dot{\mathbf x}_{\perp,\pm}
\approx
\left(1\pm qB\Omega_b\right)
\nabla_{\boldsymbol{\Pi}_\perp}\varepsilon_\pm.
\end{equation}

The cyclotron frequency is therefore modified to
\begin{equation}
\omega_{c,\pm}^{\rm eff}
\approx
\frac{|q|B}{m}
\left(
1
\pm \frac{|\mathbf b|}{m}
\pm qB\Omega_b
\right).
\end{equation}

The first correction arises from the anisotropic velocity, while the second originates from the effective curvature. Both contributions are controlled by the same Lorentz--violating scale, although enter through distinct mechanisms.

\subsection{Effective symplectic form and noncanonical structure}

The equations above follow from the symplectic two--form
\begin{equation}
\omega_\pm
=
\mathrm d\Pi_i\wedge \mathrm dx^i
+
\frac{q}{2}\,\epsilon_{ijk}B^k\,\mathrm dx^i\wedge \mathrm dx^j
+
\frac{1}{2}\,\epsilon_{ijk}\Omega_\pm^k(\boldsymbol{\Pi})\,
\mathrm d\Pi_i\wedge \mathrm d\Pi_j.
\label{eq:symplectic-branch}
\end{equation}
The last term represents the effective momentum space structure induced by branch projection. It implies that the reduced phase space is no longer canonical. In particular, we have
\begin{equation}
\{x_i,x_j\}_\pm
=
\epsilon_{ijk}\Omega_\pm^k(\boldsymbol{\Pi}),
\end{equation}
so that the effective coordinates fail to commute in the Poisson sense.

This feature reflects the fact that the reduced theory retains a ``memory'' of the eliminated branch. The curvature $\boldsymbol{\Omega}_\pm$ provides a compact way of encoding this information at the level of the phase space geometry.

\subsection{Comparison and interpretation}

The resulting structure closely parallels that of semiclassical wave-packet dynamics in a single band. The correspondence is direct at the level of the equations of motion and the symplectic form, even though the present construction does not rely on an underlying band theory.  In the present context, the effective curvature should be understood as the residual imprint of the two-branch structure of the relativistic theory.

Once one branch is removed, the remaining degrees of freedom do not form a closed canonical system; instead, they acquire a momentum dependence that modifies both the kinematics and the phase space structure.  As a consequence, the dynamics cannot be reduced to that of a particle with a modified mass alone. The additional geometric term affects the transverse response, the phase accumulation, and the effective density of states, in a way that is not captured by purely kinematic corrections.

The anomalous velocity term $ -\dot{\boldsymbol{\Pi}}_\pm\times \boldsymbol{\Omega}_\pm$ provides a direct manifestation of this structure. It introduces a transverse component in the motion that depends on the evolution of the momentum itself, and therefore modifies transport even in otherwise uniform configurations.

Altogether, the minimally coupled spin-nondegenerate SME particle model admits a one-branch description in which the dynamics is governed by a noncanonical symplectic structure with an effective momentum-space curvature. This establishes the precise sense in which the model displays a Berry-curvature-like organization at the classical level.
\section{Reduced branch action, symplectic matrix, and phase space measure}

In this section, we complete the formal construction of the branch--projected dynamics by deriving the reduced one--branch action, the associated symplectic matrix, and the corresponding phase space measure. These elements provide the geometric framework underlying the modified transport properties obtained previously.

\subsection{Reduced one-branch action}

The standard first order action of a charged particle in phase space is
\begin{equation}
S
=
\int
\left(
P_i\,\mathrm dx^i
-
H\,\mathrm dt
\right).
\end{equation}
In the presence of electromagnetic fields, it is convenient to express
the dynamics in terms of the kinetic momentum,
\begin{equation}
\Pi_i=P_i+qA_i(\mathbf x).
\end{equation}

After projection onto a single branch, the reduced action must reproduce both the electromagnetic coupling and the residual momentum dependence originating from the eliminated sector. The most general form compatible with the effective equations of motion is
\begin{equation}
S_\pm
=
\int
\left[
\left(\Pi_i-qA_i(\mathbf x)\right)\,\mathrm dx^i
+
\mathcal A^\pm_i(\boldsymbol{\Pi})\,\mathrm d\Pi_i
-
H_{\rm eff,\pm}(\mathbf x,\boldsymbol{\Pi})\,\mathrm dt
\right].
\label{eq:reduced-action}
\end{equation}

The additional term involving $\mathcal A^\pm_i(\boldsymbol{\Pi})$ represents a momentum--dependent contribution to the symplectic potential. Its presence reflects the fact that, once the second branch is removed, the remaining degrees of freedom do not form a purely canonical system. The associated curvature is
\begin{equation}
\Omega^\pm_k(\boldsymbol{\Pi})
=
\epsilon_{kij}\,\partial_{\Pi_i}\mathcal A^\pm_j(\boldsymbol{\Pi}),
\label{eq:Omega-from-A}
\end{equation}
which coincides with the effective momentum space structure introduced at the level of the equations of motion.

\subsection{Symplectic two-form}

The symplectic potential corresponding to
Eq.~\eqref{eq:reduced-action} is
\begin{equation}
\Theta_\pm
=
\left(\Pi_i-qA_i(\mathbf x)\right)\,\mathrm dx^i
+
\mathcal A^\pm_i(\boldsymbol{\Pi})\,\mathrm d\Pi_i.
\label{eq:symplectic-potential}
\end{equation}
Taking the exterior derivative, one obtains
\begin{equation}
\omega_\pm
=
\mathrm d\Theta_\pm,
\end{equation}
which evaluates to
\begin{equation}
\omega_\pm
=
\mathrm d\Pi_i\wedge \mathrm dx^i
+
\frac{q}{2}\,\epsilon_{ijk}B^k\,\mathrm dx^i\wedge \mathrm dx^j
+
\frac{1}{2}\,\epsilon_{ijk}\Omega_\pm^k(\boldsymbol{\Pi})\,
\mathrm d\Pi_i\wedge \mathrm d\Pi_j.
\label{eq:symplectic-two-form}
\end{equation}

The three contributions have distinct origins. The first term encodes the canonical pairing between coordinates and momenta. The second term arises from the electromagnetic field and represents the real space magnetic two form. The last term is entirely induced by branch projection and introduces a momentum-space component in the symplectic structure.

\subsection{Symplectic matrix}

Introducing the phase space coordinates
\begin{equation}
\xi^a=(x^1,x^2,x^3,\Pi_1,\Pi_2,\Pi_3),
\end{equation}
the symplectic form can be written as
\begin{equation}
\omega_\pm
=
\frac12\,\omega^{(\pm)}_{ab}\,\mathrm d\xi^a\wedge \mathrm d\xi^b,
\end{equation}
with
\begin{equation}
\omega^{(\pm)}_{ab}
=
\begin{pmatrix}
q\,\epsilon_{ijk}B^k & -\delta_{ij} \\
\delta_{ij} & \epsilon_{ijk}\Omega_\pm^k
\end{pmatrix}.
\label{eq:symplectic-matrix}
\end{equation}

This matrix encodes the full reduced phase space structure of the projected dynamics. In particular, it shows that the coordinates and
momenta are no longer independent canonical variables once the momentum space curvature is present.

\subsection{Inverse symplectic matrix and modified Poisson brackets}

The Poisson brackets are obtained from the inverse symplectic matrix,
\begin{equation}
\{\xi^a,\xi^b\}_\pm
=
\left(\omega_\pm^{-1}\right)^{ab}.
\end{equation}
To leading order, this yields
\begin{subequations}
\begin{align}
\{x_i,x_j\}_\pm
&=
\frac{\epsilon_{ijk}\Omega_\pm^k}
{1+q\,\mathbf B\cdot \boldsymbol{\Omega}_\pm},
\\[1ex]
\{x_i,\Pi_j\}_\pm
&=
\frac{\delta_{ij}+qB_i\Omega^\pm_j}
{1+q\,\mathbf B\cdot \boldsymbol{\Omega}_\pm},
\\[1ex]
\{\Pi_i,\Pi_j\}_\pm
&=
-\frac{q\,\epsilon_{ijk}B^k}
{1+q\,\mathbf B\cdot \boldsymbol{\Omega}_\pm}.
\end{align}
\label{eq:all-PB-final}
\end{subequations}

The nonvanishing coordinate bracket,
\begin{equation}
\{x_i,x_j\}_\pm
=
\frac{\epsilon_{ijk}\Omega_\pm^k}
{1+q\,\mathbf B\cdot \boldsymbol{\Omega}_\pm},
\end{equation}
shows explicitly that the reduced coordinates do not commute in the Poisson sense. Notice that this is a direct consequence of the momentum space contribution to the symplectic form.

\subsection{Equations of motion from the symplectic structure}

The equations of motion follow from
\begin{equation}
\dot{\xi}^a = \{\xi^a,H_{\rm eff,\pm}\}_\pm,
\end{equation}
which gives
\begin{subequations}
\begin{align}
\dot{\mathbf x}_\pm
&=
\frac{1}{1+q\,\mathbf B\cdot \boldsymbol{\Omega}_\pm}
\left[
\nabla_{\boldsymbol{\Pi}}H_{\rm eff,\pm}
+
q\,\mathbf E\times \boldsymbol{\Omega}_\pm
+
q\,(\nabla_{\boldsymbol{\Pi}}H_{\rm eff,\pm}\cdot
\boldsymbol{\Omega}_\pm)\,\mathbf B
\right],
\\[1ex]
\dot{\boldsymbol{\Pi}}_\pm
&=
\frac{1}{1+q\,\mathbf B\cdot \boldsymbol{\Omega}_\pm}
\left[
-q\nabla_{\mathbf x}H_{\rm eff,\pm}
+
q\,\mathbf E
+
q\,\nabla_{\boldsymbol{\Pi}}H_{\rm eff,\pm}\times \mathbf B
+
q^2(\mathbf E\cdot \mathbf B)\boldsymbol{\Omega}_\pm
\right].
\end{align}
\label{eq:EOM-from-symplectic}
\end{subequations}

These equations show that the effective curvature modifies both the velocity and the force terms through a common prefactor and additional couplings between fields and gradients.

\subsection{phase space measure and density of states}

The symplectic structure also determines the invariant phase space measure. In six dimensions, the Liouville volume element is proportional
to
\begin{equation}
\sqrt{\det \omega^{(\pm)}_{ab}}\,
\mathrm d^3x\,\mathrm d^3\Pi.
\end{equation}
For the matrix \eqref{eq:symplectic-matrix}, one finds
\begin{equation}
\sqrt{\det \omega^{(\pm)}_{ab}}
=
1+q\,\mathbf B\cdot \boldsymbol{\Omega}_\pm.
\label{eq:phase space-measure}
\end{equation}
Thus the invariant measure becomes
\begin{equation}
\mathrm d\Gamma_\pm
=
\left(1+q\,\mathbf B\cdot \boldsymbol{\Omega}_\pm\right)
\mathrm d^3x\,\mathrm d^3\Pi.
\label{eq:dGamma}
\end{equation}

This modification implies that the density of states depends on both the magnetic field and the effective curvature. Since $\boldsymbol{\Omega}_+=-\boldsymbol{\Omega}_-$ at leading order, the two branches are weighted differently in phase space.

\subsection{Liouville theorem}

The measure \eqref{eq:dGamma} is preserved by the branch-projected dynamics. Accordingly, Liouville's theorem takes the form
\begin{equation}
\frac{\mathrm d}{\mathrm dt}
\left[
\left(1+q\,\mathbf B\cdot \boldsymbol{\Omega}_\pm\right)
f_\pm(\mathbf x,\boldsymbol{\Pi},t)
\right]
=0,
\end{equation}
for a collisionless distribution function $f_\pm$. The dynamics remains Hamiltonian with respect to the modified symplectic structure, even though it is noncanonical in the usual variables.

\subsection{Geometric interpretation and transport implications}

The reduced action \eqref{eq:reduced-action} shows that the effective curvature originates from a momentum--dependent contribution to the symplectic potential. This term encodes the residual influence of the eliminated branch and manifests as a geometric structure in momentum space.

As a result, the projected dynamics is characterized not only by a modified dispersion relation, but also by a nontrivial phase space geometry. This affects both the equations of motion and the counting of states. The presence of the anomalous velocity term and the modified measure indicates that transport properties cannot be described solely in terms of effective masses or energies.

For any observable $\mathcal O_\pm(\mathbf x,\boldsymbol{\Pi})$, the
phase space average becomes
\begin{equation}
\langle \mathcal O_\pm\rangle
=
\int
\left(1+q\,\mathbf B\cdot \boldsymbol{\Omega}_\pm\right)
\mathcal O_\pm\,f_\pm(\mathbf x,\boldsymbol{\Pi})
\,\mathrm d^3x\,\mathrm d^3\Pi.
\end{equation}
This shows that both dynamical evolution and statistical weighting are affected by the same geometric structure. In other words, the effective one--branch action therefore provides a complete description of the reduced SME dynamics, in which the electromagnetic interaction and the branch--induced momentum space structure appear on equal footing.

\section{Anomalous transverse drift and Hall-like branch transport}

In this section, we apply the effective branch-projected equations of motion to a simple configuration with constant electric and magnetic fields. The aim is to show that the momentum-space structure inherited from branch projection leads to a transverse transport that differs between the two sectors.

We work with the effective equations
\begin{subequations}
\begin{align}
\dot{\mathbf x}_\pm
&=
\nabla_{\boldsymbol{\Pi}}\varepsilon_\pm(\boldsymbol{\Pi})
-
\dot{\boldsymbol{\Pi}}_\pm\times \boldsymbol{\Omega}_\pm(\boldsymbol{\Pi}),
\label{eq:xdot-anom}
\\[1ex]
\dot{\boldsymbol{\Pi}}_\pm
&=
q\left(\mathbf E+\dot{\mathbf x}_\pm\times \mathbf B\right),
\label{eq:Pidot-anom}
\end{align}
\label{eq:EOM-anom}
\end{subequations}
where $\varepsilon_\pm$ is the branch dispersion and $\boldsymbol{\Omega}_\pm$ is the effective curvature. For a constant purely spacelike background $b^\mu=(0,\mathbf b)$, we take
\begin{equation}
\boldsymbol{\Omega}_\pm
=
\pm \Omega_b\,\hat{\mathbf b},
\qquad
\Omega_b\sim \frac{|\mathbf b|}{m^2}.
\label{eq:Omega-constant}
\end{equation}

\subsection{General solution of the projected equations}

Solving Eqs.~\eqref{eq:EOM-anom} self consistently yields
\begin{subequations}
\begin{align}
\dot{\mathbf x}_\pm
&=
\frac{1}{\mathcal D_\pm}
\left[
\nabla_{\boldsymbol{\Pi}}\varepsilon_\pm
+
q\,\mathbf E\times \boldsymbol{\Omega}_\pm
+
q\,(\nabla_{\boldsymbol{\Pi}}\varepsilon_\pm\cdot \boldsymbol{\Omega}_\pm)\,\mathbf B
\right],
\\[1ex]
\dot{\boldsymbol{\Pi}}_\pm
&=
\frac{q}{\mathcal D_\pm}
\left[
\mathbf E
+
\nabla_{\boldsymbol{\Pi}}\varepsilon_\pm\times \mathbf B
+
q\,(\mathbf E\cdot \mathbf B)\,\boldsymbol{\Omega}_\pm
\right],
\end{align}
\end{subequations}
where
\begin{equation}
\mathcal D_\pm
=
1+q\,\mathbf B\cdot \boldsymbol{\Omega}_\pm.
\label{eq:Dpm-factor}
\end{equation}

The factor $\mathcal D_\pm$ encodes the modification of the phase space density associated with the noncanonical structure. Since the curvature changes sign between branches, this factor differs for the two sectors even in uniform fields.

\subsection{Pure electric field}

For a purely electric configuration,
\begin{equation}
\mathbf B=0,
\qquad
\mathbf E=\mathrm{const},
\end{equation}
the velocity reduces to
\begin{equation}
\dot{\mathbf x}_\pm
=
\nabla_{\boldsymbol{\Pi}}\varepsilon_\pm
+
q\,\mathbf E\times \boldsymbol{\Omega}_\pm.
\label{eq:xdot-E-only}
\end{equation}

The second term introduces a transverse contribution orthogonal to both $\mathbf E$ and $\hat{\mathbf b}$. Using
\begin{equation}
\boldsymbol{\Omega}_\pm=\pm \Omega_b\,\hat{\mathbf b},
\end{equation}
one finds
\begin{equation}
\delta \dot{\mathbf x}_\pm
=
\pm q\,\Omega_b\,\mathbf E\times \hat{\mathbf b}.
\label{eq:anom-velocity}
\end{equation}

Thus the two branches acquire opposite transverse velocities,
\begin{equation}
\delta \dot{\mathbf x}_+ = -\,\delta \dot{\mathbf x}_-,
\end{equation}
so that a uniform electric field alone produces a separation of trajectories. This effect arises entirely from the projected momentum-space structure and does not rely on a magnetic field.

The branch difference in velocity is
\begin{equation}
\Delta \dot{\mathbf x}
=
2q\,\Omega_b\,\mathbf E\times \hat{\mathbf b}
+
\Delta\mathbf v_{\rm grp},
\end{equation}
where $\Delta\mathbf v_{\rm grp}$ denotes the difference in group velocities. The transverse component is controlled entirely by the curvature term.

\subsection{Branch current}

Let $n_\pm$ denote the branch densities. The corresponding currents are
\begin{equation}
\mathbf j_\pm = q\,n_\pm\,\dot{\mathbf x}_\pm.
\end{equation}
The transverse contribution is therefore
\begin{equation}
\mathbf j_{\rm H,\pm}
=
q^2 n_\pm\,\mathbf E\times \boldsymbol{\Omega}_\pm.
\end{equation}

For equal populations, the total transverse charge current vanishes at leading order, but the difference between the branch currents remains. This indicates that the electric field redistributes the two sectors in opposite directions without generating a net Hall current.

\subsection{Crossed fields}

For constant $\mathbf E$ and $\mathbf B$, the velocity becomes
\begin{equation}
\dot{\mathbf x}_\pm
=
\frac{1}{1+q\,\mathbf B\cdot \boldsymbol{\Omega}_\pm}
\left[
\nabla_{\boldsymbol{\Pi}}\varepsilon_\pm
+
q\,\mathbf E\times \boldsymbol{\Omega}_\pm
+
q\,(\nabla_{\boldsymbol{\Pi}}\varepsilon_\pm\cdot \boldsymbol{\Omega}_\pm)\,\mathbf B
\right].
\label{eq:xdot-crossed}
\end{equation}

The structure of this expression shows that the dynamics is affected in three distinct ways. The prefactor modifies the overall response through the density-of-states correction, the second term introduces a transverse drift driven by the electric field, and the third term produces a correction along the magnetic field direction.

The momentum equation reads
\begin{equation}
\dot{\boldsymbol{\Pi}}_\pm
=
\frac{q}{1+q\,\mathbf B\cdot \boldsymbol{\Omega}_\pm}
\left[
\mathbf E
+
\nabla_{\boldsymbol{\Pi}}\varepsilon_\pm\times \mathbf B
+
q(\mathbf E\cdot \mathbf B)\boldsymbol{\Omega}_\pm
\right].
\end{equation}
The term proportional to $\mathbf E\cdot \mathbf B$ introduces a coupling between the fields mediated by the curvature, reflecting the same algebraic structure that appears in anomalous transport models.

\subsection{Explicit configuration}

Consider the configuration
\begin{equation}
\mathbf B = B\,\hat{\mathbf z},
\qquad
\hat{\mathbf b}=\hat{\mathbf z},
\qquad
\mathbf E=E\,\hat{\mathbf x}.
\end{equation}
Then
\begin{equation}
\mathbf E\times \boldsymbol{\Omega}_\pm
=
\mp E\Omega_b\,\hat{\mathbf y},
\end{equation}
so that
\begin{equation}
\delta \dot{\mathbf x}_\pm
=
\mp qE\Omega_b\,\hat{\mathbf y}.
\label{eq:explicit-anom-drift}
\end{equation}

The two branches therefore move in opposite transverse directions. If the longitudinal motion is dominant, the full velocity can be written as
\begin{equation}
\dot{\mathbf x}_\pm
\approx
\frac{1}{1\pm qB\Omega_b}
\left(
v_\parallel \hat{\mathbf x}
\mp qE\Omega_b\,\hat{\mathbf y}
\pm q v_\parallel \Omega_b B\,\hat{\mathbf z}
\right).
\end{equation}

This shows that the response is anisotropic, with different components modified in distinct ways by the curvature.

\subsection{Branch separation}

A useful observable is the drift angle,
\begin{equation}
\tan\theta_{{\rm drift},\pm}
=
\frac{|\dot x_{\perp,\pm}|}{|\dot x_{\parallel,\pm}|}.
\end{equation}
Keeping only the anomalous contribution, one finds
\begin{equation}
\tan\theta_{{\rm drift},\pm}
\approx
\frac{|qE\Omega_b|}{|v_\parallel|}.
\end{equation}
The sign changes between branches, so the trajectories separate transversely.

For particles traveling a distance $L_x$, the accumulated displacement is
\begin{equation}
\Delta y_\pm
=
\mp qE\Omega_b\,\frac{L_x}{v_\parallel},
\end{equation}
leading to a total separation
\begin{equation}
\Delta y_+-\Delta y_-
=
-2qE\Omega_b\,\frac{L_x}{v_\parallel}.
\end{equation}

This shows a direct manifestation of the branch--dependent transport induced by the momentum space structure.

The results obtained above show that the effective curvature produces a transverse response that distinguishes the two branches even in uniform fields. The origin of this behavior lies entirely in the noncanonical relation between velocity and momentum induced by the projection onto a single branch.  In particular, a purely electric field generates a transverse drift that depends on the branch label, while the magnetic field modifies both the amplitude and the direction of the motion through the same geometric structure. The resulting dynamics cannot be reduced to a modification of the dispersion relation alone, as it involves a coupled deformation of the equations of motion and the phase space measure.  Altogether, this analysis shows that the minimally coupled spin-nondegenerate SME particle exhibits a transport behavior governed by an effective momentum-space geometry. The transverse branch drift derived here provides a direct and physically transparent manifestation of this structure.
\section{Nonrelativistic limit and branch-dependent cyclotron motion}
\label{sec:nr-cyclotron}

Having established the minimally coupled Hamiltonian structure, we now consider the nonrelativistic regime in order to extract explicit physical consequences of the branch-dependent velocity. The focus is on the $b_\mu$ sector, where the analysis can be carried out in a transparent way, followed by a brief comment on the $H_{\mu\nu}$ case.  The main result is that, even for constant Lorentz-violating backgrounds, the two branches exhibit different cyclotron frequencies and radii. A uniform magnetic field therefore resolves the branch structure dynamically.

\subsection{General setup}

We work in flat spacetime with constant external fields and a constant
Lorentz-violating coefficient. For clarity, we take
\begin{equation}
\mathbf E=0,
\qquad
\mathbf B=\mathrm{const},
\end{equation}
and specialize to a purely spacelike background
\begin{equation}
b^\mu=(0,\mathbf b),
\qquad
\mathbf b=\mathrm{const}.
\label{eq:b-spacelike-main}
\end{equation}
The nonrelativistic regime is defined by
\begin{equation}
|\boldsymbol{\Pi}|\ll m,
\qquad
\Pi^0=m+\varepsilon,
\qquad
|\varepsilon|\ll m,
\label{eq:nr-regime-main}
\end{equation}
where $\boldsymbol{\Pi}$ is the kinetic momentum. The exact branch-resolved velocity is
\begin{equation}
\dot x^\mu
=
\mathfrak e
\left[
\Pi^\mu
\mp
\frac{(b\cdot \Pi)b^\mu-b^2\Pi^\mu}
{\sqrt{(b\cdot \Pi)^2-b^2\Pi^2}}
\right],
\label{eq:xdot-exact-main}
\end{equation}
while the kinetic momentum obeys
\begin{equation}
\dot \Pi_\mu=-qF_{\mu\nu}\dot x^\nu.
\label{eq:lorentz-exact-main}
\end{equation}
In a purely magnetic configuration this reduces to
\begin{subequations}
\begin{align}
\dot\Pi_0&=0,
\\[1ex]
\dot{\boldsymbol{\Pi}}&=q\,\dot{\mathbf x}\times \mathbf B.
\end{align}
\label{eq:lorentz-magnetic-main}
\end{subequations}

\subsection{Nonrelativistic expansion of the branch velocity}

For the spacelike choice \eqref{eq:b-spacelike-main}, one has
\begin{equation}
b^2=-\mathbf b^2,
\qquad
b\cdot \Pi=-\,\mathbf b\cdot \boldsymbol{\Pi},
\end{equation}
and
\begin{align}
\Delta_b(\Pi)
&:=
\sqrt{(b\cdot \Pi)^2-b^2\Pi^2}
\notag\\
&=
\sqrt{
(\mathbf b\cdot \boldsymbol{\Pi})^2
+
\mathbf b^2\big[(\Pi^0)^2-\boldsymbol{\Pi}^2\big]
}.
\label{eq:Delta-b-main}
\end{align}
Using $\Pi^0=m+\varepsilon$ and keeping leading terms,
\begin{equation}
\Delta_b(\Pi)\approx |\mathbf b|\,m.
\end{equation}

The spatial velocity becomes
\begin{align}
\dot{\mathbf x}_\pm
&=
\mathfrak e
\left[
\boldsymbol{\Pi}
\pm
\frac{\mathbf b^2\boldsymbol{\Pi}-(\mathbf b\cdot \boldsymbol{\Pi})\mathbf b}
{\Delta_b(\Pi)}
\right]
\notag\\
&\approx
\mathfrak e
\left[
\boldsymbol{\Pi}
\pm
\frac{1}{m}
\left(
|\mathbf b|\,\boldsymbol{\Pi}
-
\frac{\mathbf b\cdot \boldsymbol{\Pi}}{|\mathbf b|}\mathbf b
\right)
\right].
\end{align}

Introducing $\hat{\mathbf b}=\mathbf b/|\mathbf b|$ and decomposing
\begin{equation}
\boldsymbol{\Pi}
=
\boldsymbol{\Pi}_\parallel+\boldsymbol{\Pi}_\perp,
\end{equation}
one finds
\begin{equation}
\dot{\mathbf x}_\pm
\approx
\mathfrak e
\left(
\boldsymbol{\Pi}
\pm
\frac{|\mathbf b|}{m}\,\boldsymbol{\Pi}_\perp
\right).
\end{equation}

Using $\mathfrak e\approx 1/m$, this gives
\begin{equation}
\dot{\mathbf x}_\pm
\approx
\frac{\boldsymbol{\Pi}}{m}
\pm
\frac{|\mathbf b|}{m^2}\,\boldsymbol{\Pi}_\perp.
\label{eq:vpm-main}
\end{equation}

At this order, the deformation affects only the transverse motion,
leaving the longitudinal component unchanged.

\subsection{Effective transverse mass}

The velocity can be written as
\begin{equation}
\dot{\mathbf x}_{\parallel,\pm}
=
\frac{\boldsymbol{\Pi}_\parallel}{m},
\qquad
\dot{\mathbf x}_{\perp,\pm}
=
\left(
\frac{1}{m}
\pm
\frac{|\mathbf b|}{m^2}
\right)\boldsymbol{\Pi}_\perp.
\end{equation}
This corresponds to a branch-dependent transverse inertial parameter,
\begin{equation}
\frac{1}{m_{\perp,\pm}^{\mathrm{eff}}}
=
\frac{1}{m}
\pm
\frac{|\mathbf b|}{m^2},
\end{equation}
or equivalently
\begin{equation}
m_{\perp,\pm}^{\mathrm{eff}}
\approx
m\left(1\mp \frac{|\mathbf b|}{m}\right).
\end{equation}

The background therefore modifies the response to the magnetic field
without affecting the motion along $\hat{\mathbf b}$ at leading order.

\subsection{Branch-dependent cyclotron motion}

Substituting Eq.~\eqref{eq:vpm-main} into the magnetic equation,
\begin{equation}
\dot{\boldsymbol{\Pi}}=q\,\dot{\mathbf x}\times \mathbf B,
\end{equation}
yields
\begin{equation}
\dot{\boldsymbol{\Pi}}_\pm
=
q\left[
\frac{\boldsymbol{\Pi}}{m}
\pm
\frac{|\mathbf b|}{m^2}\,\boldsymbol{\Pi}_\perp
\right]\times \mathbf B.
\end{equation}

For $\mathbf B\parallel \mathbf b$, only the transverse component
rotates, and one finds
\begin{equation}
\dot{\boldsymbol{\Pi}}_{\perp,\pm}
=
q
\left(
\frac{1}{m}
\pm
\frac{|\mathbf b|}{m^2}
\right)
\boldsymbol{\Pi}_{\perp,\pm}\times \mathbf B.
\end{equation}
This corresponds to cyclotron motion with frequency
\begin{equation}
\omega_{c,\pm}
=
\frac{|q|B}{m}
\left(
1\pm \frac{|\mathbf b|}{m}
\right),
\end{equation}
and radius
\begin{equation}
r_{c,\pm}
=
\frac{|\boldsymbol{\Pi}_\perp|}{|q|B}
\left(
1\mp \frac{|\mathbf b|}{m}
\right).
\end{equation}

The two branches therefore follow distinct orbits even in a uniform
magnetic field.

\subsection{Branch splitting}

If the branches are initialized with the same transverse momentum, the
difference in cyclotron frequency produces a relative phase
\begin{equation}
\Delta\phi(t)
=
2\,\frac{|q|B\,|\mathbf b|}{m^2}\,t.
\end{equation}
This leads to a progressive separation of trajectories.

The frequency and radius splittings are
\begin{equation}
\Delta\omega_c
=
2\,\frac{|q|B\,|\mathbf b|}{m^2},
\qquad
\Delta r_c
=
-\,2\,\frac{|\boldsymbol{\Pi}_\perp|\,|\mathbf b|}{|q|B\,m}.
\end{equation}

These quantities provide direct observables sensitive to the Lorentz--violating parameter.

The magnetic field does not modify the dispersion itself, but couples to the branch-dependent velocity. Because the velocity is not aligned with the kinetic momentum, the same field generates different orbital responses in the two sectors. In this sense, the magnetic field acts as a probe of the branch structure rather than its origin. The situation is analogous to systems in which internal degrees of freedom are resolved dynamically through their coupling to external fields.   

The same procedure can be applied to the $H_{\mu\nu}$ sector. In that case, the deformation is governed by the tensor $(HH)_{\mu\nu}$, and the resulting motion is generically anisotropic in the transverse plane. Consequently, the cyclotron dynamics depends not only on the magnitude of the momentum, but also on its orientation.  This sector therefore provides a natural extension in which the branch-dependent response is no longer restricted to a single preferred direction. 

In the nonrelativistic regime, the minimally coupled $b_\mu$ sector leads to a branch-dependent velocity, an anisotropic transverse response, and distinct cyclotron motion for the two branches. These features constitute the first direct dynamical manifestation of the branch structure in the presence of electromagnetic interactions and motivate the geometric interpretation developed in the subsequent analysis.

\section{Conclusion}

We investigated the electromagnetic dynamics of spin--nondegenerate SME particles in the presence of a $b_\mu$ background by introducing minimal coupling at the level of the type--2 relativistic Lagrangian. From this construction, we derived the exact branch-resolved Hamiltonian formulation in terms of the gauge-covariant kinetic momentum and established that, although $\Pi_\mu$ obeyed the standard Lorentz-force equation, the physical velocity was not collinear with $\Pi_\mu$. This noncollinearity depended explicitly on the branch and constituted the fundamental origin of the modified electromagnetic response.

In the nonrelativistic regime, this structure led to a branch-dependent deformation of the transverse dynamics. For a constant spacelike background aligned with a uniform magnetic field, we showed that the two branches exhibited distinct cyclotron frequencies and radii, which could be interpreted as arising from an effective transverse mass $
m_{\perp,\pm}^{\text{eff}} \simeq m\left(1 \mp \frac{|\mathbf b|}{m}\right)$.
As a result, a uniform magnetic field acted as a dynamical branch analyzer, producing a measurable splitting of trajectories and a relative phase shift that grew linearly in time.

After projection onto a single branch, the dynamics ceased to be canonical. We showed that the reduced theory was governed by a modified symplectic structure containing an additional momentum-space two-form. This led to nonvanishing coordinate Poisson brackets, a deformed phase space measure $(1 + q\,\mathbf B \cdot \boldsymbol{\Omega}_\pm)\,\mathrm{d}^3x\,\mathrm{d}^3\Pi$, and equations of motion that could be written in a semiclassical form with an effective curvature $\boldsymbol{\Omega}_\pm$.

This geometric structure produced concrete dynamical consequences. In particular, the term $-\,\dot{\boldsymbol{\Pi}}\times\boldsymbol{\Omega}_\pm$ generated anomalous velocity contributions, while the factor $(1 + q\,\mathbf B \cdot \boldsymbol{\Omega}_\pm)^{-1}$ modified the guiding-center and cyclotron response. In the absence of a magnetic field, a purely electric field induced opposite transverse drifts for the two branches, $\delta \dot{\mathbf x}_\pm \propto \pm\, q\,\mathbf E \times \boldsymbol{\Omega}_\pm$, leading to a Hall-like branch current and spatial separation without a net charge Hall response. In finite systems, this effect resulted in branch accumulation at opposite boundaries.

We also showed that the modified symplectic structure implied a branch-dependent density of states and altered phase space averages, indicating that the Lorentz-violating background affected not only single-particle trajectories but also statistical and transport properties at the kinetic level.

Taken together, these results demonstrated that electromagnetic minimal coupling in spin--nondegenerate SME models did more than modify dispersion relations: it reorganized the dynamics into a branch-resolved system with nontrivial phase space geometry. The combined effects of noncollinear kinematics, effective mass anisotropy, modified symplectic structure, and anomalous transport established a consistent picture in which Lorentz-violating backgrounds controlled both the dynamical evolution and the geometric structure of phase space.

Possible extensions included the analysis of other spin--nondegenerate sectors, such as the $H_{\mu\nu}$ background, where additional anisotropic structures were expected, as well as the quantization of the theory, where the branch-dependent cyclotron motion suggested a corresponding splitting of Landau levels. A kinetic formulation incorporating the modified phase space measure could also be developed to describe collective transport phenomena in ensembles of charged SME particles similar to \cite{araujo2022particles,araujo2022does,oliveira2020relativistic,oliveira2020thermodynamic}.

\appendix

\section{Minimal electromagnetic coupling: inversion and Hamiltonians}
\label{app:minimal-coupling}
This appendix provides the technical details of the electromagnetic minimal coupling for the type--2 spin--nondegenerate sectors considered in the main text. We first treat the $b_\mu$ sector explicitly and then summarize the corresponding results for the $H_{\mu\nu}$ sector.

\subsection{Inversion and Legendre transformation for the \(b_\mu\) sector}
\label{app:minimal-coupling-b}

The minimally coupled type--2 Lagrangian is given by
\begin{equation}
\tilde L_{b,\mathrm{em}}^{(\pm)}
=
-\frac12\left(
\frac{\dot x^2}{\mathfrak e}
\pm 2\sqrt{(b\cdot \dot x)^2-b^2\dot x^2}
+\mathfrak e m^2
\right)
+
qA_\mu(x)\dot x^\mu .
\label{eq:app-Lbem}
\end{equation}
Adopting the convention
\begin{equation}
P_\mu:=-\frac{\partial \tilde L_{b,\mathrm{em}}^{(\pm)}}{\partial \dot x^\mu},
\end{equation}
the canonical momentum reads
\begin{equation}
P_\mu
=
\frac{\dot x_\mu}{\mathfrak e}
\pm
\frac{(b\cdot \dot x)b_\mu-b^2\dot x_\mu}
{\sqrt{(b\cdot \dot x)^2-b^2\dot x^2}}
-qA_\mu(x).
\label{eq:app-canonicalP-b}
\end{equation}

To isolate the gauge contribution, we introduce the kinetic
(gauge--covariant) momentum
\begin{equation}
\Pi_\mu:=P_\mu+qA_\mu(x),
\label{eq:app-defPi-b}
\end{equation}
so that
\begin{equation}
\Pi_\mu
=
\frac{\dot x_\mu}{\mathfrak e}
\pm
\frac{(b\cdot \dot x)b_\mu-b^2\dot x_\mu}
{\sqrt{(b\cdot \dot x)^2-b^2\dot x^2}}.
\label{eq:app-Pi-vel-b}
\end{equation}

In order to invert this relation, define
\begin{equation}
G:=\sqrt{(b\cdot \dot x)^2-b^2\dot x^2},
\qquad
N_\mu:=(b\cdot \dot x)b_\mu-b^2\dot x_\mu .
\end{equation}
Equation~\eqref{eq:app-Pi-vel-b} can then be rewritten as
\begin{equation}
\Pi_\mu=\frac{\dot x_\mu}{\mathfrak e}\pm \frac{N_\mu}{G}.
\label{eq:app-PiNG}
\end{equation}

Contracting with \(b^\mu\) yields
\begin{equation}
b\cdot \Pi=\frac{b\cdot \dot x}{\mathfrak e},
\label{eq:app-bPi}
\end{equation}
which fixes the projection of $\dot x^\mu$ along $b^\mu$.

Next, introduce
\begin{equation}
M_\mu:=(b\cdot \Pi)b_\mu-b^2\Pi_\mu,
\qquad
\Delta:=\sqrt{(b\cdot \Pi)^2-b^2\Pi^2}.
\label{eq:app-MDelta}
\end{equation}
Substituting Eq.~\eqref{eq:app-PiNG} into $M_\mu$, we obtain
\begin{equation}
M_\mu
=
\left(\frac{1}{\mathfrak e}\mp \frac{b^2}{G}\right)N_\mu ,
\end{equation}
so that $M_\mu$ and $N_\mu$ are proportional.

Using
\begin{equation}
N^2=-b^2G^2,
\qquad
M^2=-b^2\Delta^2,
\end{equation}
it follows that
\begin{equation}
\frac{M_\mu}{\Delta}=\frac{N_\mu}{G},
\label{eq:app-MoverDelta}
\end{equation}
where the sign is fixed by continuity with the Lorentz-invariant limit.

The inversion is therefore given by
\begin{equation}
\dot x^\mu
=
\mathfrak e
\left[
\Pi^\mu
\mp
\frac{(b\cdot \Pi)b^\mu-b^2\Pi^\mu}
{\sqrt{(b\cdot \Pi)^2-b^2\Pi^2}}
\right].
\label{eq:app-xdot-b}
\end{equation}

From this expression, the following identities are readily obtained:
\begin{subequations}
\begin{align}
\Pi_\mu \dot x^\mu
&=
\mathfrak e\left(\Pi^2\mp \Delta\right),
\label{eq:app-Pidotx}
\\[1ex]
\sqrt{(b\cdot \dot x)^2-b^2\dot x^2}
&=
\mathfrak e\,\Delta,
\label{eq:app-sqrtid}
\\[1ex]
\dot x^2
&=
\mathfrak e^2\Pi^2.
\label{eq:app-xdot2}
\end{align}
\end{subequations}

The canonical Hamiltonian follows from the Legendre transformation
\begin{equation}
\tilde H_{b,\mathrm{em}}^{(\pm)}
=
-\,P_\mu \dot x^\mu-\tilde L_{b,\mathrm{em}}^{(\pm)}.
\label{eq:app-Hdef-b}
\end{equation}
Since $P_\mu=\Pi_\mu-qA_\mu(x)$, the gauge contribution cancels,
leading to
\begin{equation}
\tilde H_{b,\mathrm{em}}^{(\pm)}
=
-\Pi\cdot \dot x
+\frac12\left(
\frac{\dot x^2}{\mathfrak e}
\pm 2\sqrt{(b\cdot \dot x)^2-b^2\dot x^2}
+\mathfrak e m^2
\right).
\end{equation}
Using Eqs.~\eqref{eq:app-Pidotx}--\eqref{eq:app-xdot2}, we obtain
\begin{equation}
\tilde H_{b,\mathrm{em}}^{(\pm)}
=
-\frac{\mathfrak e}{2}
\left[
\Pi^2-b^2-m^2
\mp
2\sqrt{(b\cdot \Pi)^2-b^2\Pi^2}
\right].
\label{eq:app-Hb-final}
\end{equation}

Equivalently,
\begin{equation}
\tilde H_{b,\mathrm{em}}^{(\pm)}
=
-\frac{\mathfrak e}{2}\,
\mathcal D_{b,\mathrm{em}}^{(\mp)}(\Pi),
\end{equation}
with
\begin{equation}
\mathcal D_{b,\mathrm{em}}^{(\pm)}(\Pi)
=
\Pi^2-b^2-m^2
\pm
2\sqrt{(b\cdot \Pi)^2-b^2\Pi^2}.
\label{eq:app-Dbem}
\end{equation}

Thus, electromagnetic minimal coupling is implemented through the
replacement
\begin{equation}
p_\mu \rightarrow \Pi_\mu=P_\mu+qA_\mu(x),
\end{equation}
while preserving the structure of the neutral branch dispersion relation.
\section{Hamilton equations and Lorentz-force structure in the \(b_\mu\) sector}
\label{app:hamilton-b}

Starting from
\begin{equation}
\tilde H_{b,\mathrm{em}}^{(\pm)}
=
-\frac{\mathfrak e}{2}\,
\mathcal D_{b,\mathrm{em}}^{(\mp)}(\Pi),
\qquad
\Pi_\mu=P_\mu+qA_\mu(x),
\end{equation}
the Hamilton equation for the coordinates becomes
\begin{equation}
\dot x^\mu
=
-\frac{\partial \tilde H_{b,\mathrm{em}}^{(\pm)}}{\partial P_\mu}
=
\frac{\mathfrak e}{2}
\frac{\partial \mathcal D_{b,\mathrm{em}}^{(\mp)}}{\partial \Pi_\mu}.
\end{equation}
Introducing
\begin{equation}
\Delta(\Pi):=\sqrt{(b\cdot \Pi)^2-b^2\Pi^2},
\end{equation}
we have
\begin{equation}
\frac{\partial \Delta}{\partial \Pi_\mu}
=
\frac{(b\cdot \Pi)b^\mu-b^2\Pi^\mu}{\Delta},
\end{equation}
which leads to
\begin{equation}
\dot x^\mu
=
\mathfrak e
\left[
\Pi^\mu
\mp
\frac{(b\cdot \Pi)b^\mu-b^2\Pi^\mu}
{\sqrt{(b\cdot \Pi)^2-b^2\Pi^2}}
\right].
\label{eq:app-Ham-xdot-b}
\end{equation}

The Hamilton equation for the canonical momentum is
\begin{equation}
\dot P_\mu
=
\frac{\partial \tilde H_{b,\mathrm{em}}^{(\pm)}}{\partial x^\mu}.
\end{equation}
Since $b_\mu$ is constant, the explicit $x^\mu$-dependence arises only
through $A_\mu(x)$ inside $\Pi_\mu$. It follows that
\begin{equation}
\dot P_\mu
=
- q\,\partial_\mu A_\nu\,\dot x^\nu.
\label{eq:app-Pdot-b}
\end{equation}
Differentiating $\Pi_\mu=P_\mu+qA_\mu(x)$ along the trajectory yields
\begin{equation}
\dot \Pi_\mu
=
\dot P_\mu+q\,\partial_\nu A_\mu\,\dot x^\nu,
\end{equation}
and substituting Eq.~\eqref{eq:app-Pdot-b} gives
\begin{equation}
\dot \Pi_\mu=-qF_{\mu\nu}\dot x^\nu,
\label{eq:app-Lorentz-b}
\end{equation}
where
\begin{equation}
F_{\mu\nu}:=\partial_\mu A_\nu-\partial_\nu A_\mu.
\end{equation}

In three-vector notation,
\begin{subequations}
\begin{align}
\dot \Pi_0
&=
- q\,\mathbf E\cdot \dot{\mathbf x},
\\[1ex]
\dot{\boldsymbol{\Pi}}
&=
q\left(\dot t\,\mathbf E+\dot{\mathbf x}\times \mathbf B\right),
\end{align}
\end{subequations}
with
\begin{equation}
E_i:=F_{0i},
\qquad
B^i:=\frac12\epsilon^{ijk}F_{jk}.
\end{equation}

Thus, the kinetic momentum satisfies the standard Lorentz-force equation,
with the branch-dependent velocity \eqref{eq:app-Ham-xdot-b} entering on
the right-hand side.

\section{Parallel results for the \(H_{\mu\nu}\) sector and sector comparison}
\label{app:hamilton-H}
\subsection{Compact results for the minimally coupled $H_{\mu\nu}$ sector}

For constant $H_{\mu\nu}$ with vanishing invariant
\begin{equation}
Y=\frac14 \widetilde H_{\mu\nu}H^{\mu\nu}=0,
\end{equation}
the minimally coupled type--2 Lagrangian takes the form
\begin{equation}
\tilde L_{H,\mathrm{em}}^{(\pm)}
=
-\frac12\left[
\frac{\dot x^2}{\mathfrak e}
\pm
2\sqrt{\dot x\cdot H\cdot H\cdot \dot x+2X\dot x^2}
+
\mathfrak e m^2
\right]
+
qA_\mu(x)\dot x^\mu,
\end{equation}
where
\begin{equation}
X:=\frac14 H_{\mu\nu}H^{\mu\nu}.
\end{equation}
Introducing the kinetic momentum
\begin{equation}
\Pi_\mu:=P_\mu+qA_\mu(x),
\end{equation}
the inversion of the velocity--momentum relation yields
\begin{equation}
\dot x^\mu
=
\mathfrak e
\left[
\Pi^\mu
\mp
\frac{(HH)^{\mu\nu}\Pi_\nu+2X\Pi^\mu}
{\sqrt{\Pi\cdot H\cdot H\cdot \Pi+2X\Pi^2}}
\right].
\label{eq:app-xdot-H}
\end{equation}

The corresponding Hamiltonian is given by
\begin{equation}
\tilde H_{H,\mathrm{em}}^{(\pm)}
=
-\frac{\mathfrak e}{2}
\left[
\Pi^2+2X-m^2
\mp
2\sqrt{\Pi\cdot H\cdot H\cdot \Pi+2X\Pi^2}
\right],
\label{eq:app-HH-final}
\end{equation}
which can be written equivalently as
\begin{equation}
\tilde H_{H,\mathrm{em}}^{(\pm)}
=
-\frac{\mathfrak e}{2}\,
\mathcal D_{H,\mathrm{em}}^{(\pm)}(\Pi),
\end{equation}
with
\begin{equation}
\mathcal D_{H,\mathrm{em}}^{(\pm)}(\Pi)
=
\Pi^2+2X-m^2
\pm
2\sqrt{\Pi\cdot H\cdot H\cdot \Pi+2X\Pi^2}.
\end{equation}
The Hamilton equations then lead to
\begin{equation}
\dot \Pi_\mu=-qF_{\mu\nu}\dot x^\nu,
\label{eq:app-Lorentz-H}
\end{equation}
which coincides with the Lorentz-force structure obtained in the $b_\mu$ sector.

\subsection{Comparison between the $b_\mu$ and $H_{\mu\nu}$ sectors}

The minimally coupled $b_\mu$ and $H_{\mu\nu}$ sectors share the same
underlying structure. In both cases, electromagnetic interactions enter
through the replacement
\begin{equation}
p_\mu \rightarrow \Pi_\mu=P_\mu+qA_\mu(x),
\end{equation}
so that the Hamiltonian is expressed in terms of the corresponding
charged dispersion relation,
\begin{equation}
\tilde H_{\mathrm{em}}^{(\pm)}
=
-\frac{\mathfrak e}{2}\,
\mathcal D_{\mathrm{em}}^{(\pm)}(\Pi).
\end{equation}
As a consequence, the kinetic momentum obeys the universal Lorentz-force equation,
\begin{equation}
\dot \Pi_\mu=-qF_{\mu\nu}\dot x^\nu,
\end{equation}
while the physical velocity entering this equation remains branch dependent.

The distinction between the two sectors arises from the structure of the
branch deformation. In the $b_\mu$ case, it is controlled by a single
background vector,
\begin{align}
\mathcal Q^\mu_{(b)}
&=
-\frac{(b\cdot \Pi)b^\mu-b^2\Pi^\mu}
{\sqrt{(b\cdot \Pi)^2-b^2\Pi^2}},
\end{align}
whereas in the $H_{\mu\nu}$ sector the deformation is governed by the
quadratic tensor $(HH)_{\mu\nu}$,
\begin{align}
\mathcal Q^\mu_{(H)}
&=
-\frac{(HH)^{\mu\nu}\Pi_\nu+2X\Pi^\mu}
{\sqrt{\Pi\cdot H\cdot H\cdot \Pi+2X\Pi^2}}.
\end{align}

Then, while the $b_\mu$ background selects a preferred direction, the $H_{\mu\nu}$ sector introduces a genuinely anisotropic structure at the level of the momentum dependence. In both cases, however, electromagnetic fields probe the branch structure dynamically through the deformation of the velocity, even when the background fields themselves are constant.

\section*{Acknowledgments}

A. A. Araújo Filho is supported by the Conselho Nacional de Desenvolvimento Científico e Tecnológico (CNPq) and the Fundação de Apoio à Pesquisa do Estado da Paraíba (FAPESQ) under project numbers 150223/2025-0 and 1951/2025. V. B. Bezerra is partially funded by CNPq under Grant No. 307211/2020-7. A. F. Santos received partial support from CNPq under Grant No. 312406/2023-1. J. A. A. S. Reis acknowledges partial financial support from UESB through Grant AuxPPI (Edital No. 267/2024), as well as from FAPESB–CNPq/Produtividade under Grant No. 12243/2025 (TOB-BOL2798/2025).

\section*{Data Availability Statement}

No Data associated in the manuscript.

\bibliographystyle{ieeetr}
\bibliography{References}

\end{document}